\documentclass[aps, prl, showpacs, twocolumn, amsfonts, amsmath, amssymb, superscriptaddress, floatfix]{revtex4}

\usepackage{graphicx}
\usepackage{dcolumn}
\usepackage{bm}

\begin{document}

\title{Shortcut to adiabaticity for an interacting Bose-Einstein condensate}

\author{Jean-Fran\c{c}ois Schaff}
\affiliation{Universit\'e de Nice-Sophia Antipolis, Institut Non Lin\'eaire de Nice, CNRS, 1361 route des Lucioles, F-06560 Valbonne, France}

\author{Xiao-Li Song}
\affiliation{Universit\'e de Nice-Sophia Antipolis, Institut Non Lin\'eaire de Nice, CNRS, 1361 route des Lucioles, F-06560 Valbonne, France}

\author{Pablo Capuzzi}
\affiliation{Universidad de Buenos Aires, FCEN, Departamento de Fisica and Instituto de Fisica de Buenos Aires, CONICET, Ciudad Universitaria, Pab. I C1428EGA Buenos Aires, Argentina}

\author{Patrizia Vignolo}
\affiliation{Universit\'e de Nice-Sophia Antipolis, Institut Non Lin\'eaire de Nice, CNRS, 1361 route des Lucioles, F-06560 Valbonne, France}

\author{Guillaume Labeyrie}
\email{guillaume.labeyrie@inln.cnrs.fr}
\affiliation{Universit\'e de Nice-Sophia Antipolis, Institut Non Lin\'eaire de Nice, CNRS, 1361 route des Lucioles, F-06560 Valbonne, France}


\begin{abstract}
We present an investigation of the fast decompression of a three-dimensional (3D) Bose-Einstein condensate (BEC) at finite temperature using an engineered trajectory for the harmonic trapping potential. Taking advantage of the scaling invariance properties of the time-dependent Gross-Pitaevskii equation, we exhibit a solution yielding a final state identical to that obtained through a perfectly adiabatic transformation, in a much shorter time. Experimentally, we perform a large trap decompression and displacement within a time comparable to the final radial trapping period. By simultaneously monitoring the BEC and the non-condensed fraction, we demonstrate that our specific trap trajectory is valid both for a quantum interacting many-body system and a classical ensemble of non-interacting particles.

\end{abstract}

\maketitle

Quantum adiabatic transformations~\cite{Born1928,Kato1950}, in which the system's parameters are varied slowly enough such that no transition between instantaneous eigenstates occur, play a central role in physics. For instance, schemes based on adiabatic passage have been proposed to prepare non classical states~\cite{Parkins1993,Cirac1994} or to produce new strongly correlated states~\cite{Sorensen2010}. Quantum adiabatic computation is attracting a lot of attention~\cite{Peng2008, Rezakhani2009}. Adiabatic transformations using various time-dependent potentials are routinely performed in experiments on ultracold gases. However adiabatic transformations are typically slow~\cite{Comparat2009}, while practical applications and experimental constraints such as finite lifetime or coherence time~\cite{Zurek2003} require faster processes.

This contradiction motivated the search for rapid schemes reproducing or approaching ideal adiabatic transitions. ``Exact'' methods~\cite{Berry2009,Chen2010a}, here referred to as \emph{shortcuts to adiabaticity}, yield a final state strictly identical to that obtained via an adiabatic transformation, while other approaches~\cite{Schulz2006, Hohenester2007, Vasilev2009, Mundt2009} use minimization techniques to optimize the transition to a target state. Among the former, some strategies require introducing additional terms in the Hamiltonian~\cite{Berry2009,Chen2010b}, others consist in engineering the time dependence of the parameters to avoid unwanted transitions~\cite{Muga2010}. In spite of the large theoretical literature, few experiments were conducted on classical systems~\cite{Couvert2008, Wang2010, Schaff2010}, and even fewer in the quantum regime~\cite{DeChiara2008}.

\begin{figure}
\begin{center}
\resizebox{1.05\columnwidth}{!}{\includegraphics{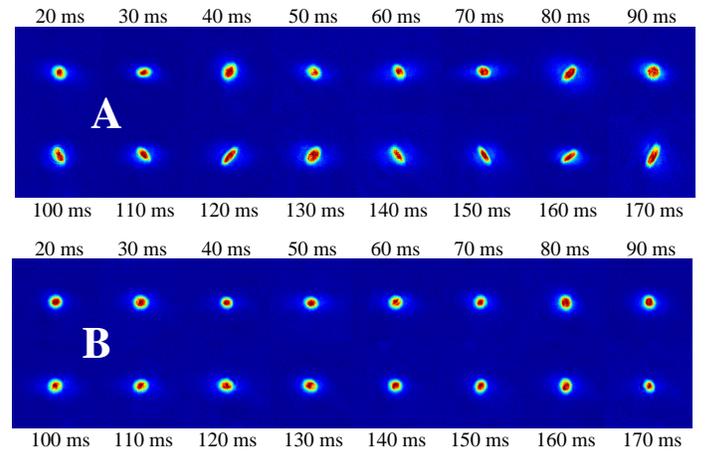}} 
\caption{Linear versus shortcut BEC decompression. We compare the time evolution of the BEC after two different decompression schemes: (\textbf{A}) a 30-ms-long linear ramp and (\textbf{B}) the shortcut trajectory (see text). The center of mass motion has been subtracted from these time of flight images for clarity.}
\label{fig1}
\end{center}
\end{figure}

In this Letter, we perform the rapid shortcut decompression of a 3D interacting BEC confined in an anisotropic harmonic trap. The trap frequencies are decreased by a factor of 9 (radially) and 3 (axially) in a time comparable to the final radial trapping period, using a trajectory based on the scaling properties of the time-dependent Gross-Pitaevskii equation (GPE) in the Thomas-Fermi (TF) limit~\cite{Kagan1996}. This shortcut trajectory leads to a final state identical (in theory) to the equilibrium state obtained via a perfectly adiabatic process. Experimentally, we demonstrate that the collective excitations~\cite{Jin1996,Mewes1996} associated with the rapid trap decompression are strongly reduced by our shortcut scheme (Fig.~\ref{fig1}), the residual excitation being due to experimental imperfections. Furthermore, we show that the trajectory is also valid for a classical ensemble of non-interacting particles, as demonstrated by monitoring the non-condensed fraction of the finite-temperature BEC. 

We describe our system by a zero-temperature BEC plus a thermal cloud, assumed to behave independently. The BEC component thus obeys the 3D GPE
\begin{equation}
  i\hbar \frac{\partial}{\partial t} \psi(\mathbf r, t)
  =
  \bigg[-\frac{\hbar^2}{2m}\nabla^2 + U(\mathbf r,t) + 
\tilde U N|\psi(\mathbf r, t)|^2
  \bigg] \psi(\mathbf r, t) \,,
  \label{GPE}
\end{equation}
where $\psi(\mathbf r, t)$ is the wave function of the condensate, $m$ the mass, $N$ the number of particles, and
$\tilde U=4\pi\hbar^2 a_s/m$ the interaction coupling constant ($a_s$ is the s-wave scattering length). The time-dependent trapping potential has a cylindrical symmetry along the horizontal axis ($y$) 
\begin{equation}
  U(\mathbf r,t) = \frac{1}{2}m\,\omega_\perp(t)^2(x^2+z^2) + \frac{1}{2} m\,\omega_\parallel^2(t)y^2 + mgz\,,
  \label{pot}
\end{equation}
with initial and final angular frequencies $\omega_{0\{\perp,\parallel\}}$ and $\omega_{f\{\perp,\parallel\}}=\omega_{0\{\perp,\parallel\}}/\gamma^2_{\perp,\parallel}$, respectively. It is worth stressing here that decreasing the trap frequencies not only decompresses the BEC but also translates the harmonic potential minimum vertically by $\Delta z = -g (1/\omega_{0\perp}^2-1/\omega_{f\perp}^2)$. The objective is to engineer a trajectory $\omega_{\perp,\parallel}(t)$ connecting the \textit{equilibrium states} in the initial and final potentials. We stress that the BEC is not at equilibrium at any time during the trajectory, but only at $t=0$ and $t=t_f$. 

Equation (\ref{GPE}) is invariant under the scaling and translational transformation
\begin{equation}
 \psi(\mathbf r, t)=(b_\perp^2b_\parallel)^{-1/2}\chi(\pmb{\rho},\tau(t))
\exp[i\phi(\mathbf r, t)]
\end{equation}
with $\rho_x=x/b_\perp $, $\rho_y=y/b_\parallel$, 
$\rho_z=z/b_\perp+ga/\omega_{0\perp}^2$ and $\tau(t) = \int_0^t dt'/[b_\perp^2(t')b_\parallel(t')]$, in the following situations: either (i) in the non-interacting limit~\cite{Kagan1996,DGO2009}; or (ii) for a suitable driving of the interaction term $\tilde U$ via 
Feshbach resonances~\cite{DGO2009}; or (iii) in the Thomas-Fermi (TF) limit (i.e. neglecting the term $\nabla^2\chi$)~\cite{Castin1997}.
 
In the latter case, one can derive three coupled differential equations satisfied by the scaling 
($b_{\perp,\parallel}$) and shifting ($a$) parameters, the condition of initial and final equilibrium imposing sixteen independent boundary conditions. Our procedure to engineer $\omega_{\perp,\parallel}(t)$ is to reduce the dimensionality of the problem by looking for trajectories with a \textit{constant} $b_\parallel$ (and thus a constant axial size of the BEC), yielding a decompression with $\gamma_\perp=\gamma_\parallel^2$. In this case, $b_{\perp,\parallel}$ and $a$ should satisfy
\begin{gather}
\label{eq:bperp1}
\ddot b_\perp(t)+ b_\perp(t)\omega_\perp^2(t) = 
\omega_{0\perp}^2/b_\perp(t)^3 \\
\label{eq:bparallel1}
\omega_\parallel(t) = 
\omega_{0\parallel}/b_\perp(t)\\
\label{eq:a1}
b_\perp(t)^4\ddot a(t)+2b_\perp(t)^3
\dot b_\perp(t)\dot a(t)
+\omega_{0\perp}^2a(t)-\omega_{0\perp}^2b_\perp(t)^3=0,
\end{gather}
for which we can exploit the procedure outlined in \cite{Schaff2010}. Since Eqs.~(\ref{eq:bperp1}) and (\ref{eq:a1}) are identical to those obtained for a thermal cloud~\cite{Schaff2010}, the shortcut trajectory exhibited here is valid \textit{both} for the BEC (in all directions) and for the thermal fraction in the radial directions. Indeed, since Eq.~(\ref{eq:bparallel1}) does not yield a fixed axial size for a thermal cloud, this shortcut trajectory will not work for the axial direction in the thermal case.

\begin{figure}
\begin{center}
\resizebox{1.0\columnwidth}{!}{\includegraphics{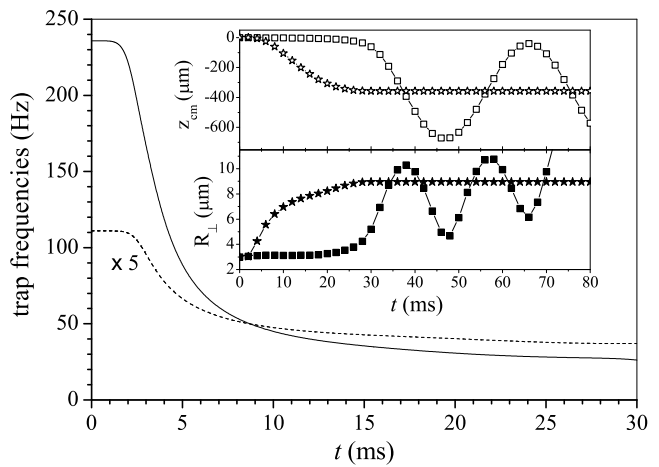}} 
\caption{Shortcut BEC decompression in 30 ms. We plot the shortcut trajectories $\omega_{\perp}(t)/2\pi$ (solid) and $\omega_{\parallel}(t)/2\pi$ ($\times 5$, dashed). The insert compares the subsequent evolution of the BEC's center of mass (open symbols) and radial size (solid symbols) for the shortcut (stars) and linear (squares) decompressions (GPE simulation).}
\label{fig2}
\end{center}
\end{figure}

Our ultracold $^{87}$Rb atoms are trapped in a quadrupole-Ioffe-configuration (QUIC) magnetic trap~\cite{Hansch1998}. In its compressed initial state this trap is anisotropic with radial and axial frequencies $\omega_{0\perp}/2\pi = 235.8$ Hz and $\omega_{0\parallel}/2\pi = 22.2$ Hz respectively. We can tune these independently by adjusting the QUIC current (affects both $\omega_\perp$ and $\omega_\parallel$) and the current running through a pair of Helmoltz coils aligned along $y$ (affects only $\omega_\perp$)~\cite{Foot2002}. In the present paper, we perform a 30-ms-long radial decompression of the trap by a factor of 9, yielding a final radial frequency $\omega_{f\perp}/2\pi = 26.2$ Hz. Because of the condition $\gamma_\perp=\gamma_\parallel^2$ imposed above, the axial frequency is reduced by a factor of 3 to a final value $\omega_{f\parallel}/2\pi = 7.4$ Hz. If performed adiabatically, this decompression reduces the chemical potential by a factor of 9. The corresponding shortcut frequency trajectories are represented on Fig.~\ref{fig2}. The insert shows the expected evolution of the vertical center of mass position of the BEC and its radial size as obtained from GPE numerical simulations. For the shortcut trajectory (stars), these quantities remain stationary after the end of the decompression indicating that an equilibrium final state is reached. For comparison, we plot (squares) the same quantities for a linear ramp of same duration, yielding large dipolar and breathing oscillations~\cite{Dalfovo1999}.

The experimental procedure is as follows. We first produce a BEC by RF evaporation in the compressed trap. In the experimental runs presented here, the condensed fraction (N $= 1.3 \times 10^5$) represents $60\%$ of the total number of atoms. The initial temperature, inferred from the size of the non-condensed fraction after time of flight, is $T_0 = 130$ nK. We then apply a decompression sequence, hold the ultracold cloud for a certain time $t_h$ in the decompressed trap,  then release it and monitor the cloud's parameters after a 28 ms time of flight via absorption imaging. This time of flight is close to the critical time ($\approx 30$ ms) where the aspect ratio of the decompressed BEC inverts, which explains its isotropic aspect on Fig.~\ref{fig1}. By varying $t_h$, we characterize the magnitude of the various modes excited by the decompression process. To extract quantitative estimates, we fit the 2D column density profiles by a two-component distribution allowing for different angles (see discussion below) for the BEC (TF profile) and thermal fraction (Gaussian profile). The fit results are averaged over three different images taken in the same conditions. 
Throughout this paper we will compare three different decompression schemes: an abrupt jump from the initial to final frequencies, a linear ramp of duration 30 ms, and the shortcut trajectory depicted on Fig.~\ref{fig2}.

\begin{figure}
\begin{center}
\resizebox{0.95\columnwidth}{!}{\includegraphics{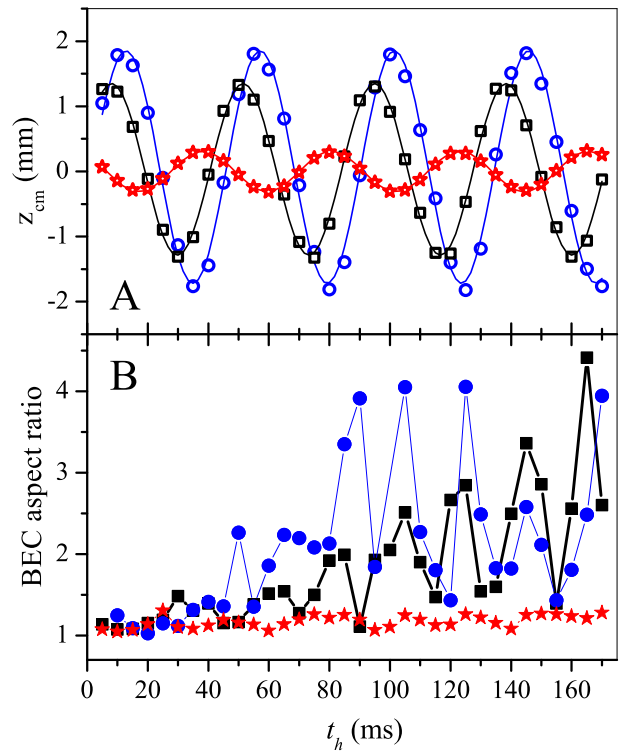}} 
\caption{(Color online) Decompression-induced excitations. We report the temporal evolution of (\textbf{A}) the center of mass position and (\textbf{B}) the aspect ratio of the BEC after three different decompression schemes: an abrupt decompression (open circles); a 30 ms linear ramp (squares); the 30ms shortcut trajectory (stars). All measurement are performed after a 28 ms-long time of flight.}
\label{fig3}
\end{center}
\end{figure}

Fig.~\ref{fig1} illustrates the efficiency of our shortcut method. The vertical axis on the figure corresponds to the direction of gravity ($z$), and the horizontal one to the axial direction ($y$). The field of view of each image is 576 $\mu$m $\times$ 576 $\mu$m, the indicated value corresponding to the time $t_h$ spent in the decompressed trap. The center of mass motion has been subtracted from these images for clarity (see Fig.~\ref{fig3}(\textbf{A})). (\textbf{A}) corresponds to a 30-ms-long linear decompression while (\textbf{B}) is obtained with the shortcut trajectory. Qualitatively, we observe that in the first case the BEC undergoes large deformations characterized by an oscillatory behavior of its aspect ratio. More unexpectedly, the BEC is also seen to oscillate \emph{angularly} about the horizontal axis, reflecting the excitation of a scissors mode~\cite{Stringari1999,Marago2000}. This ``parasitic'' excitation will be discussed at the end of the paper. 
Applying the shortcut scheme results in a drastic suppression of the BEC deformations (Fig.~\ref{fig1}(\textbf{B})). Indeed, the BEC remains almost stationary throughout the whole 170 ms range after decompression, very close to the targeted equilibrium state.

Fig.~\ref{fig3}(\textbf{A}) shows the oscillation occurring at $\omega_{f\perp}$ of the cloud's center of mass position along the vertical (after a 28 ms time of flight). This dipole mode is excited by the large vertical displacement (357 $\mu$m) of the trap center. The circles, squares and stars correspond to the abrupt, linear and shortcut decompressions respectively. The lines are sinusoidal fits. The shortcut decompression reduces the amplitude of the dipole mode by a factor of 6 and 4.3 when compared to the abrupt and linear schemes respectively. The corresponding peak vertical velocity is 1.8 recoil. The origin of this residual dipole oscillation will be discussed at the end of the paper. Note that the center of mass oscillation is totally uncoupled from the internal degrees of freedom of the BEC, even in the presence of interactions~\cite{Dalfovo1999}.

\begin{figure}
\begin{center}
\resizebox{0.95\columnwidth}{!}{\includegraphics{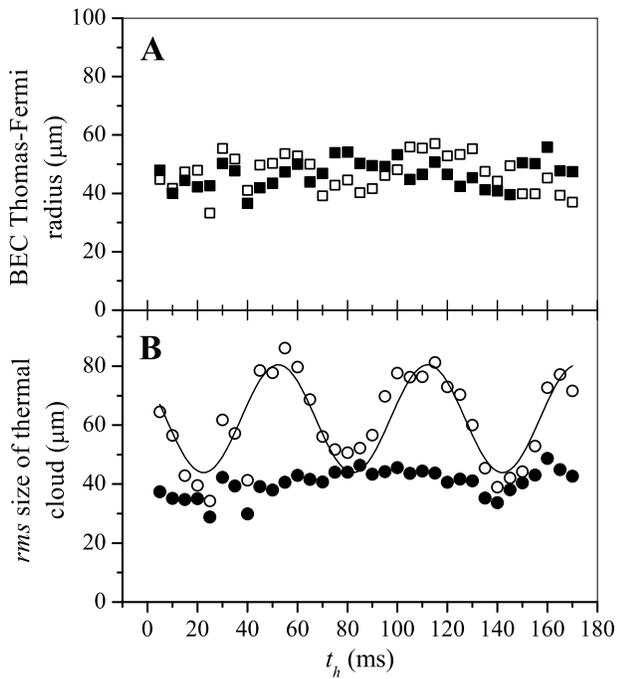}} 
\caption{BEC versus thermal cloud decompression. We plot the sizes of the BEC (\textbf{A}) and thermal component (\textbf{B}) versus $t_h$ for the shortcut trajectory. The filled and empty symbols correspond to the radial (vertical) and axial directions respectively.}
\label{fig4}
\end{center}
\end{figure}

We report on Fig.~\ref{fig3}(\textbf{B}) the evolution of the aspect ratio of the BEC (again after a 28 ms time of flight). For the abrupt and linear schemes, it presents strong oscillations at a frequency of 47 Hz which is consistent with a radial breathing mode of frequency $\approx 2\omega_{f\perp}/2\pi$~\cite{Mewes1996, Chevy2002, Dalfovo1999}. A Fourier transform analysis of the BEC size oscillations also reveals the presence of an axial breathing oscillation at 12.5 Hz ($\approx \sqrt{5/2}~\omega_{f\parallel}/2\pi$~\cite{Dalfovo1999}). The specific shape of the aspect ratio fluctuations on Fig.~\ref{fig3}(\textbf{B}) results from the interplay between the various modes and depends on the decompression trajectory as confirmed by GPE simulations. Using the shortcut trajectory (stars on Fig.~\ref{fig3}(\textbf{B})) strongly inhibits these breathing-like excitations: the standard deviation of the aspect ratio variations versus $t_h$ is reduced by a factor of 12 and 10 when compared to the abrupt and linear decompressions respectively. After the shortcut decompression and the 28-ms-long expansion, the BEC has an average TF radius of 46.8 $\mu$m (close to the GPE prediction of 43 $\mu$m) with a standard deviation below 12\% (see Fig.~\ref{fig4}(\textbf{A})).

Fig.~\ref{fig4} illustrates the impact of the shortcut decompression on the thermal fraction. Here, as mentioned before, the shortcut trajectory is expected to work only for the radial directions. Indeed, we observe on Fig.~\ref{fig4}(\textbf{B}) a breathing oscillation at $2\times\omega_{f\parallel}/2\pi$ of the thermal cloud along the axial direction (open circles on lower panel), while its size along the vertical (filled circles) remains stationary, corresponding to a temperature $T_f = 22$ nK (a factor of 5.9 below the initial temperature). In the case of the linear ramp, we observe breathing oscillations along both axial and radial dimensions. For comparison, we show on (\textbf{A}) the sizes of the BEC along both axis, which are stationary since the shortcut trajectory is valid both radially and axially.

We now discuss the various imperfections of the experiment, which limit the performances of our shortcut decompression and result in a final state different from the targeted equilibrium one. A first problem is the mismatch between the theoretical and experimental trap frequency trajectories. For instance, we measure from the dipole oscillation $\omega_{f\perp}/2\pi = 23.6$ Hz which is $10\%$ below the targeted value of 26.2 Hz. Since the cloud's dynamics is quite sensitive to the final stage of the decompression, this can account for the residual dipole oscillation of Fig.~\ref{fig3}(\textbf{A}). Another possible issue is the trap anharmonicity, which is enhanced in the decompressed trap due to gravity. The impact of the anharmonicity could be reduced by using a smoother trajectory~\cite{Schaff2010, Couvert2008}. The trap's non-ideal geometry is also responsible for the excitation of the scissors mode observed on Fig.~\ref{fig1}. In Ref.~\cite{Marago2000}, an angular velocity was imparted to the BEC by suddenly tilting the trap, and the subsequent monochromatic angular oscillation at $\sqrt{\omega_\perp^2+\omega_\parallel^2}$ was used as an evidence for superfluidity. Here, the angular momentum is communicated to the cloud during the trap decompression because the trap eigenaxes tilt slightly as the trap center moves downwards due to gravity. This results in an angle of $3^{\circ}$ between the axial directions of the initial and final traps in the vertical plane. Note that the large amplitude of the scissors oscillation on Fig.~\ref{fig1} is due to the magnification effect of the time of flight~\cite{Edwards2002, Modugno2003}. In-situ measurement show an amplitude of the scissors oscillation compatible with the trap tilt angle of $3^{\circ}$.

In conclusion, we presented in this paper a method to perform shortcut-to-adiabaticity transformations on a 3D interacting BEC, using a specifically designed parameter trajectory for the harmonic trapping potential. The performances could be further improved using better-controlled potentials such as in optical traps or lattices, where time-dependent manipulations are also easier and faster. Very short transition times could in principle be achieved by transiently applying negative (i.e. expelling) curvatures~\cite{Chen2010a}. Further work may include the direct comparison with other methods such as ``bang-bang''~\cite{Salamon2009} or optimal control techniques. More general shortcut solutions will also be searched for, and applied to other dimensionalities or non-harmonic potentials~\cite{Lewis1982}. These fast transition methods are not restricted to cold atom manipulation, and can be readily adapted to topics as diverse as e.g. macroscopic resonator cooling~\cite{Li2010}, temporal~\cite{Chen2010b} and spatial~\cite{Jong2010} coherent population transfer, or quantum computation~\cite{Rezakhani2009}.

\acknowledgments{This work was supported by CNRS and Universit\'e de Nice-Sophia Antipolis. We also acknowledge financial support from R\'egion PACA, F\'ed\'eration Wolfgang Doeblin, and CNRS-CONICET international cooperation grant n$^\circ$22966.}

\end{document}